\documentclass[
	aps,
	prx,
	amsmath,
	reprint,
	superscriptaddress,
	longbibliography
]{revtex4-2}
\usepackage{hyperref}
\usepackage{graphicx}
\usepackage{units}
\usepackage{upgreek}
\usepackage{color}
\usepackage{amsmath, amssymb, amsfonts}
\makeatletter
\renewcommand\subparagraph{\@startsection{subparagraph}{5}{\z@}{3.25ex\@plus1ex\@minus.2ex}{-1em}{\normalfont\normalsize\bfseries}}

\renewcommand{\@make@capt@title}[2]{\@ifx@empty\float@link{\@firstofone}{\expandafter\href\expandafter{\float@link}}{\textsf{\bfseries#1}}\@caption@fignum@sep{\textsf{#2}}}

\renewcommand\@caption@fignum@sep{\textsf{\bfseries.}}

\let\oldtheequation\theequation
\renewcommand\tagform@[1]{\maketag@@@{\ignorespaces#1\unskip\@@italiccorr}}
\renewcommand\theequation{(\oldtheequation)}
\makeatother

\usepackage{tabularx}

\begin{document}

\begin{abstract}
	Dynamical clustering represents a characteristic feature of active matter consisting of self-propelled agents that convert energy from the environment into mechanical motion. At the micron scale, typical of overdamped dynamics, particles with opposite motility block each other and show transient dynamical arrest that can induce cluster nucleation and motility-induced phase separation. However, for macroscopic agents, inertia plays a leading role, and clustering is strongly affected by bounce-back effects during collisions that could inhibit cluster growth. Here we present an experiment on the clustering of active granular particles, in which inertia can be systematically tuned. A plethora of new phenomena impacted severely by inertia is presented. Clusters display an inertia-induced transition from liquid-like to crystal-like inner structures that is accompanied by the suppression of cluster nucleation at system boundaries due to particle inertia. 
In contrast to microswimmers, where active particles wet the boundary by primarily forming clusters attached to the container walls, in an underdamped inertial active system, walls do not favor cluster formation and effectively annihilate motility-induced wetting phenomena.
\end{abstract}

\title{Dynamical clustering and wetting phenomena in inertial active matter}

\author{Lorenzo Caprini}
\email{lorenzo.caprini@hhu.de, lorenzo.caprini@uniroma1.it} 
\affiliation{Institut f\"ur Theoretische Physik II: Soft Matter, Heinrich-Heine-Universit\"at D\"usseldorf, D-40225 D\"usseldorf, Germany}
\affiliation{Department of Physics, University of Rome La Sapienza, 00185, Rome, Italy.}

\author{Davide Breoni}
\affiliation{Institut f\"ur Theoretische Physik II: Soft Matter, Heinrich-Heine-Universit\"at D\"usseldorf, D-40225 D\"usseldorf, Germany}

\author{Anton Ldov}
\affiliation{Institut f\"ur Theoretische Physik II: Soft Matter, Heinrich-Heine-Universit\"at D\"usseldorf, D-40225 D\"usseldorf, Germany}

\author{Christian Scholz}
\affiliation{Institut f\"ur Theoretische Physik II: Soft Matter, Heinrich-Heine-Universit\"at D\"usseldorf, D-40225 D\"usseldorf, Germany}

\author{Hartmut L\"owen}
\affiliation{Institut f\"ur Theoretische Physik II: Soft Matter, Heinrich-Heine-Universit\"at D\"usseldorf, D-40225 D\"usseldorf, Germany}

\date{\today}

\maketitle

\section{Introduction}

A large variety of physical systems exhibits complex cluster formation. The birth of planets, for example, is induced by dust agglomeration, galaxies organize into multi-scale clusters on astronomical length scales, while molecular clusters are the building blocks for nanoparticles.
This phenomenon is pivotal on the mesoscale, for example in systems of overdamped colloidal particles which attract each other because of van der Waals forces~\cite{stradner2004equilibrium, lu2006fluids}. In this framework, clustering is also central in the vibrant research arena of active matter~\cite{marchetti2013hydrodynamics, Elgeti2015}, consisting of self-propelling agents that convert energy from the environment into directed motion.
Typical examples are cells, bacteria, and active colloids at the micron scale, or animals, robots, and active granular particles at the macroscopic scale. 
These systems usually display collective phenomena, such as flocking and swarming, typical of birds~\cite{cavagna2014bird}, fish~\cite{jhawar2020noise}, insects~\cite{cavagna2017dynamic}, but also bacteria~\cite{zhang2010collective}, Quincke rollers~\cite{bricard2013emergence}, and active granular particles with elongated shapes~\cite{kudrolli2008swarming, kumar2014flocking} or mass asymmetry~\cite{deseigne2010collective}.
In this case, active particles move coherently, self-organizing in a polarized state.

\begin{figure*}[t]
	\includegraphics[width=\textwidth]{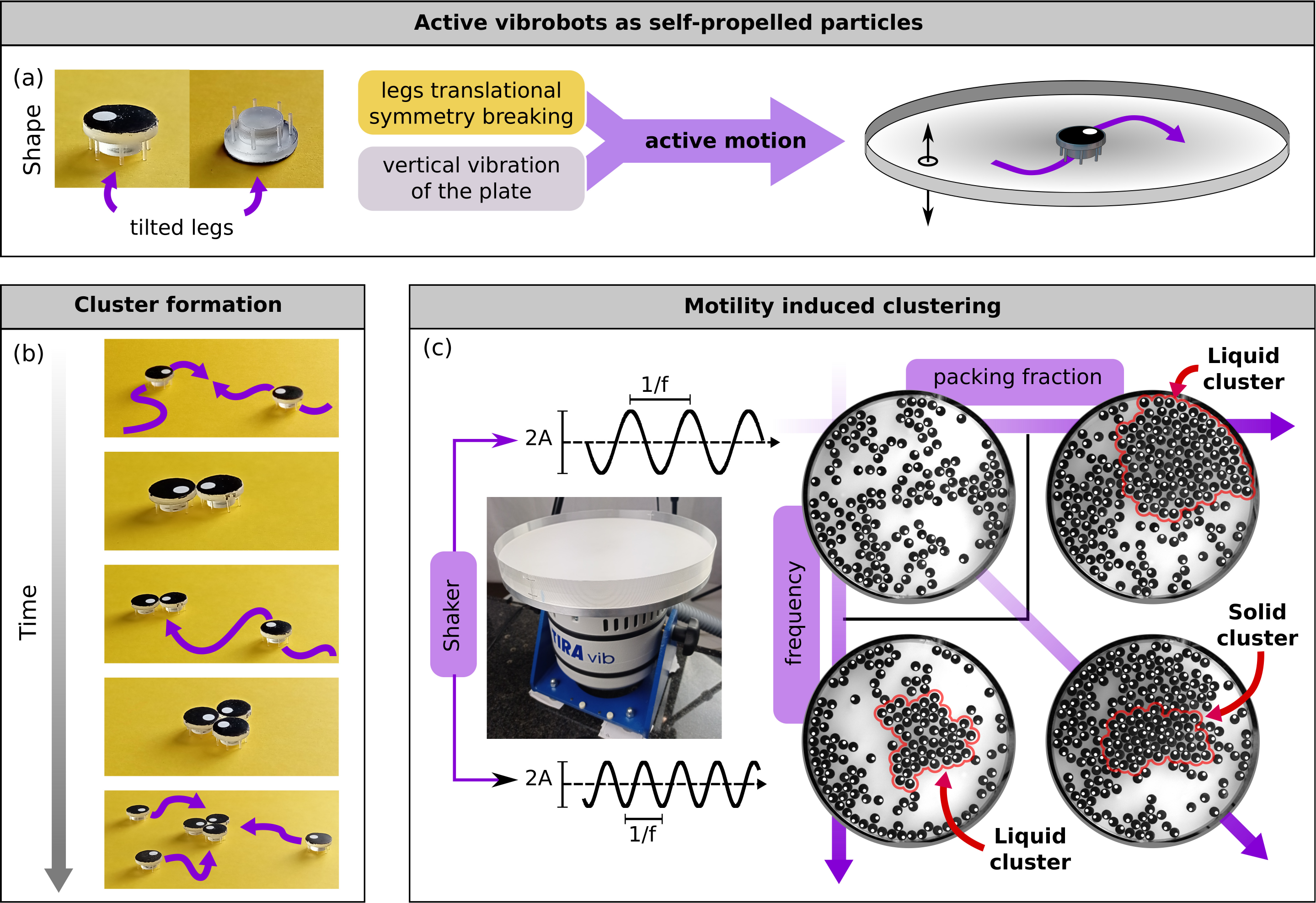}
	\caption{
		\label{Fig:Fig_start}
		\textbf{Motility induced clustering in active granular particles.}
		(a)~Photos of an active particle (top and bottom view). The seven legs attached to the body are tilted by an angle of $4$ degrees.
		The direction of propulsion is marked by the white spot on the top of the particle.
		(b)~Sketch of a typical time evolution of a growing cluster consisting of active particles.
		Particle trajectories are marked by violet lines. At the initial time, two particles are stuck after a collision. A third particle approaches the pair and gets stuck, giving rise to a cluster consisting of three particles. Successively, other particles approach the cluster, giving rise to a nucleation event.
		(c)~Images from the experimental recordings with packing fractions $\phi=0.3$ and $\phi=0.45$ and frequencies $f=\unit[90]{Hz}$ and $f=\unit[150]{Hz}$.
		At low packing fraction and small frequency of the shaker, the system does not show cohesive structures. By increasing $f$ and/or $\phi$ large clusters with liquid-like or solid-like structures are observed.
	}
\end{figure*}

Crowded active systems may also display unpolarized and highly dynamical clustering phenomena, typically at high densities and large particle motilities.
Cluster nucleation is generated by particles persistently moving in opposite directions and on occasion blocking each other's motion, causing transient arrest.
When particles are stuck in a pair or a triplet like that, the nucleation of a cluster can be observed if surrounding particles approach the group (as illustrated in Fig.~\ref{Fig:Fig_start}~{(a)-(b)}).
This may lead to a phase coexistence between a cluster and a dilute phase, termed motility-induced phase separation~\cite{Cates2015}.
Over the last decade, motility-induced clustering phenomena, or simply dynamical clustering, have been largely explored for active colloids swimming on a two-dimensional substrate where the motion of the individual active particles is overdamped due to the embedding fluid.
In this case, colloidal experiments~\cite{Palacci2013, ginot2018aggregation, Buttinoni2013, ginot2015nonequilibrium, van2019interrupted, geyer2019freezing} have confirmed cluster formation and illuminated its dependence on density and motility. Meanwhile, a large body of simulations and theoretical works have been performed for overdamped self-propelled particles~\cite{Fily2012, klamser2018thermodynamic, caprini2020spontaneous, caporusso2020motility, grossmann2020particle, siebers2023exploiting} to characterize the basic mechanisms inducing clustering and develop coarse-grained theories.
However, for macroscopic systems, particle inertia plays a leading role: two inertial active particles continuously bounce back during a collision.
This mechanism hampers the mutual arrest and therefore the cluster formation.
Though inertia is key for macroscopic particles, its effect on clustering has not been systematically explored in active systems and experiments are currently missing.
\begin{figure*}
	\includegraphics[width=\textwidth]{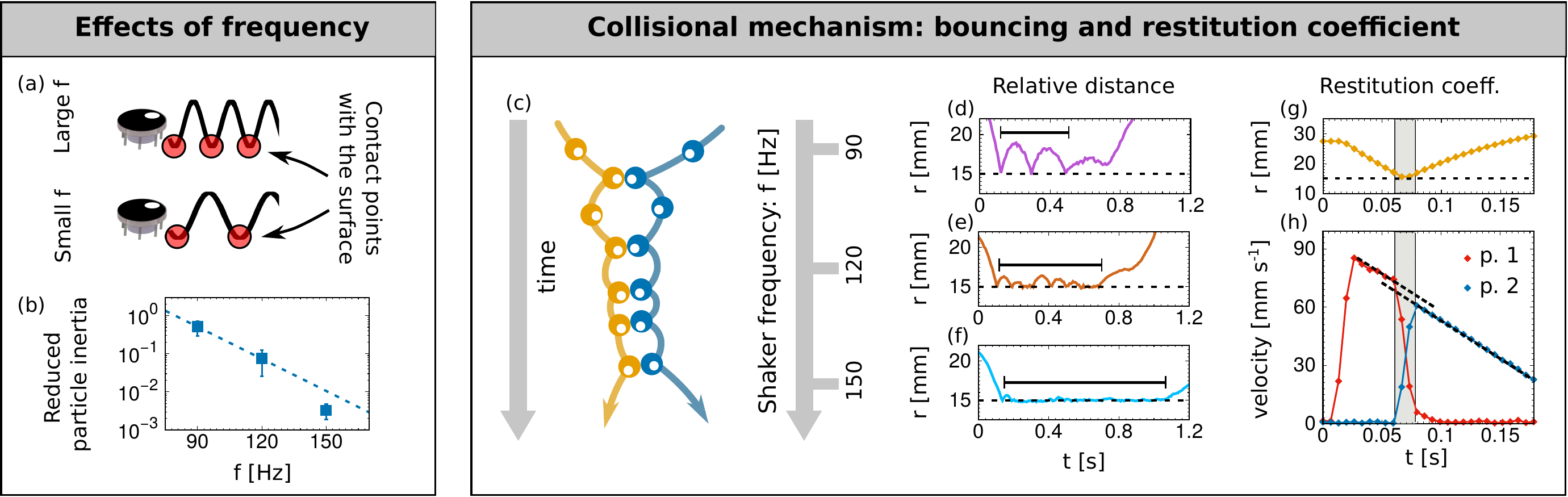}
	\caption{
		\label{Fig:Fig_inertia}
		\textbf{Frequency-tuned inertia.}
		(a)~Schematic effect of shaker frequency on the vertical motion of a vibrobot: with higher frequency, legs and surface touch more often, increasing friction and reducing inertia. 
		(b)~Reduced particle inertia, $M=m/(\gamma\tau)$, defined as the ratio between the inertial time (mass over friction) and the persistence time, as a function of the shaker's frequency for a single active vibrobot (see Appendix~\ref{sec:experimental}).
		(c)~Illustration of two colliding particles' trajectories, showing the bouncing mechanism.
		(d)-(f)~Relative distance between two colliding particles for different values of the shaker's frequency: $f=\unit[90]{Hz}$ (d), $f=\unit[120]{Hz}$ (e), $f=\unit[150]{Hz}$ (f). Dashed black lines in each graph mark two particle radii, i{.}e{.} the collision distance. Vertical horizontal bars denote the duration of a typical collision which increases with frequency.
		(g)~Relative distance, $r(t)$, between two passive particles colliding after a kick (see SM, Sec.~S1).
		(h)~Corresponding velocity of the kicked particle (p. 1, red) and of the second particle (p. 2, blue). Black dashed lines in this graph are extrapolations of the collisional velocities. In (g) and (h), gray regions mark the time interval wherein the collision occurs.
		Error bars in (b) represent the standard deviation. 
	}
\end{figure*}

Here, we close this gap and perform an experiment on active granules where inertia can be systematically varied (Fig.~\ref{Fig:Fig_start}~{(a)}).
The bounce-back effect during a collision leads to a reduction of cluster formation: inertia hampers motility-induced clustering (Fig.~\ref{Fig:Fig_start}~{(c)}) and changes the phase diagram, unlike in passive equilibrium systems.
At high density, we observe a plethora of novel effects which can be modulated by inertia.
Interestingly, the inner cluster structure can be tuned to be either liquid-like or solid-like (Fig.~\ref{Fig:Fig_start}~{(c)}), in contrast to overdamped systems which typically form clusters with inner crystalline order.
Moreover, we show that wetting phenomena, i.e.\ nucleation of clusters at system boundaries, can be completely annihilated by inertia.
Our work may stimulate future theories and serve to pave some of the way towards smart active materials with tunable microstructure using inertial effects.

\section{Results}

\subsection{Experimental setup}
Our experiment features macroscopic self-propelled (active) particles called vibrobots, manufactured by 3D printing (see Fig.~\ref{Fig:Fig_start}~(a)). An electromagnetic shaker (see Fig.~\ref{Fig:Fig_start}~(c)) with excitation frequency $f$ and amplitude $A$ transfers vertical vibrations to a plate where particles are placed.
Plate vibrations induce quasi-two-dimensional dynamics of the particles since their vertical motion is negligible.

All active particles consist of a cylindrical body (see Appendix~\ref{sec:experimental} for details) to which several legs are attached.
These legs are all identically tilted with a typical angle that determines the particle polarization. This is also marked by a white spot on the particle's top side (Fig.~\ref{Fig:Fig_start}~(a)) for visual tracking.
The tilted elastic legs enable directed self-propelled motion when the system is excited by vibrations: the active particle bounces forward because of a ratcheting mechanism originating from the asymmetric collisions between legs and plate.
At small times, particles move in the polarization direction with a typical constant self-propulsion velocity.
Surface inhomogeneities and a bouncing instability produce small random reorientations, that, at long times, randomize the particle motion and determine the vibrobot persistence time. 
As a result, each self-propelled vibrobot behaves as an active Brownian particle~\cite{bechinger2016active, shaebani2020computational} with inertia~\cite{Scholz2018inertial}, due to the macroscopic size of the particles.
In addition, the cylindrical shape of the particle body is designed to reduce self-alignment mechanisms~\cite{baconnier2022selective}, i{.}e{.} the tendency of the orientation to align with the particle velocity.
This is confirmed by visual inspection of the supplementary movies 1, 2, and 3 (see also Sec.~S3 of Supplemental Material, SM). 
Further details and the intrinsic properties of the vibrobots design are reported in Appendix~\ref{sec:experimental}.

Despite the granular properties of our system, collisions between vibrobots are almost elastic, i{.}e{.} they almost conserve energy.
Evidence for this is provided by experimentally estimating the restitution coefficient in a typical collision between two particles (Fig.~\ref{Fig:Fig_inertia}~(g)-(h)). 
The time trajectories of their relative distance (Fig.~\ref{Fig:Fig_inertia}~(g)) allow us to estimate the amount of time the particles spend in quasi-contact during a collision. 
By comparing the time evolution of vibrobot speed before and after the collision (Fig.~\ref{Fig:Fig_inertia}~(h)), a value for the restitution coefficient can be estimated and reads $e\approx 0.9$ (see SM, Sec.~S1).

In our experiments, we tune the frequency and the amplitude
of the sinusoidal vibration to control vibrobot properties. We also vary the packing fraction $\Phi=N\frac{d^2}{4R^2}$, where $N$ is the number of active particles, $d$ is their diameter, and $R$ is the radius of the circular plate.
Our investigation ranges from a dilute system with $\Phi=0.15$, where particles interact rarely, up to high densities with $\Phi=0.6$.

\subsection{Tuning the effective particle inertia}
The dynamics of our active particles is characterized by frictional forces proportional to particle velocities through an effective friction coefficient~\cite{Scholz2018}.
These forces have a purely mechanical origin and, for individual particles, are induced by the friction between vibrobot legs and the surface of the plate.
By decreasing the shaker frequency $f$, we observe a slight change in the P\'eclet number (see Appendix~\ref{sec:parameters}) but, mostly, a consistent increase of the reduced particle inertia, $M=\frac m{\gamma\tau}$, i{.}e{.} the ratio between the inertial time (particle mass $m$, over effective friction coefficient, $\gamma$) and the active force reorientation time $\tau$.
Intuitively, decreasing the frequency enhances the flying time of a vibrobot in the vertical direction and thus reduces the contact time between legs and plate (Fig.~\ref{Fig:Fig_inertia}~{(a)}).
This decreases the effective friction experienced by the particle and thus increases $M$ (Fig.~\ref{Fig:Fig_inertia}~{(b)}).
The effective inertia increase manifests itself in the typical collisions between two active vibrobots:
two self-propelled particles usually collide when their polarizations point roughly toward one another.
After the impact, both particles bounce back, i{.}e{.} their relative velocities are reversed.
Then, the self-propulsion mechanism restores much of the pre-collisional velocity by pushing the particles against each other again.
This mechanism leads to several bouncing events during the collisions between active vibrobots (see the illustration in Fig.~\ref{Fig:Fig_inertia}~(c)), before random reorientation releases them from one another.
These multi-collision events have a typical duration that depends on the velocity memory and thus are determined by the inertial time $m/\gamma$.
This picture is confirmed by monitoring the relative distance, $r$, between the centers of two colliding particles.
If the reduced inertia is small (large $f$), the velocity change is almost instantaneous and $r$ is almost constant during the collision (Fig.~\ref{Fig:Fig_inertia}~(f)). This reproduces analogous results known from microscopic systems, such as colloids or bacteria that persistently push against obstacles, boundaries~\cite{vladescu2014filling, junot2022run} or other particles.
Large effective inertia (small $f$), on the other hand, induces more pronounced bouncing events with a confining wall~\cite{leoni2020surfing} or between active particles. As a result, the relative distance $r$ displays oscillations whose amplitude increases on average for smaller $f$ (Fig.~\ref{Fig:Fig_inertia}~(d)-(e)).
In short, the smaller the $f$, the larger the reduced inertia and the more pronounced bouncing is observed.

\begin{figure*}[!t]
\includegraphics[width=\textwidth]{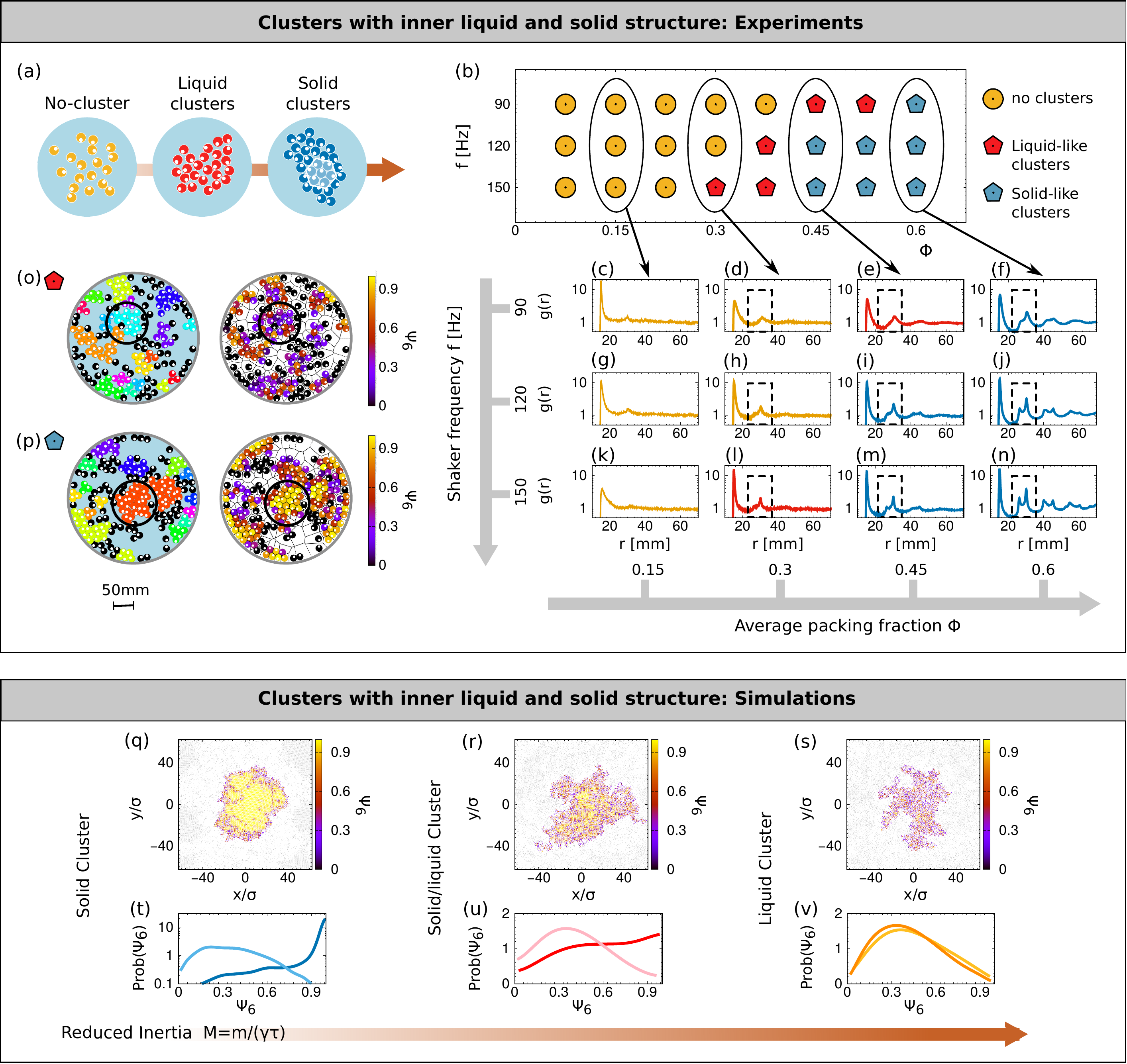}
	\caption{
		\label{Fig:Fig_phase}
		\textbf{Inner structure of clusters: from solid to liquid order}.
		(a)~Illustrations of particle clusters for increasing packing fraction $\Phi$.
		(b)~Phase diagram in the plane of packing fraction $\Phi$ and shaker frequency $f$. Yellow circles denote homogeneous phase, red pentagons motility-induced clustering with a liquid-like structure, and blue pentagons motility-induced clustering with a solid-like structure.
		(c)-(n)~Pair correlation function $g(r)$ for $f=\unit[90]{Hz}$, $\unit[120]{Hz}$, $\unit[150]{Hz}$, and packing fraction $\Phi=0.15,0.3,0.45,0.6$.
		(o)-(p)~Experimental snapshots at $\Phi=0.45$, $f=\unit[90]{Hz}$ (o) and $\Phi=0.525$, $f=\unit[120]{Hz}$ (p). In the left snapshot, particles are colored according to the cluster they belong, while, in the right snapshot, colors represent $\Psi_6$ (see Appendix~\ref{sec:psi6}, Eq.~\eqref{eq:Psi6} for the $\Psi_6$ definition). Black circles in the snapshots outline a cluster with a liquid-like (o) and a solid-like structure (p).
		(q)-(s)~Snapshot configurations obtained by simulations with number of particles $N=10^4$ and box size $L=125.3$ ($\Phi=0.5$). Colors plot the $\Psi_6$ heat map.
		(t)-(v)~Corresponding probability distribution $\text{Prob}(\Psi_6)$ calculated inside and outside the cluster. 
		All the dimensionless parameters are kept fixed (see Appendix~\ref{sec:simulations}) except the reduced particle inertia, $M$, which determines the particle inertia: (q), (t) ($M=10^{-2}$), (r), (u) ($M=5\times10^{-2}$), and (s), (v) ($M=10^{-1}$).
	}
\end{figure*}

\subsection{Inertia-induced liquid-crystal transition in the cluster}

We investigate dynamical cluster formation for several values of the average packing fraction $\Phi$ and shaker frequency $f$.
Configurations showing clustering are identified by studying the local packing fraction distribution $P(\phi)$, such that $\langle\phi\rangle=\Phi$ (see SM, Sec.~S2): The bimodality of $P(\phi)$ is used as a criterion to determine the relevance of dynamical clustering in the system since this is a signature of phase separation between a dense and a dilute phase~\cite{PhysRevLett.110.055701}.
Results are summarized in a phase diagram (Fig.~\ref{Fig:Fig_phase}~(b)) distinguishing, at first, cluster and no-cluster configurations (Fig.~\ref{Fig:Fig_phase}~(b)).
Dynamical clustering is promoted by the increase of $\Phi$, as occurs in passive particles with attractive interactions.
By contrast, our experiments prove that the clustering of active granular particles is suppressed when reduced inertia is increased (small $f$), as previously predicted in simulations~\cite{omar2023tuning}: This occurs when collisions are dominated by inertia-induced bouncing-back mechanisms which hamper arrested states.
The cluster suppression phenomenon has a pure non-equilibrium origin since, in passive equilibrium systems, inertia cannot affect the phase diagram.

Our first central result concerns an inertia-induced structural transition in the cluster, which is observed by studying the pair correlation function $g(r)$.
This analysis reveals that high values of reduced inertia (small shaker frequency) alter the structural properties inside dense clusters by inducing a solid-liquid transition where hexagonal order is suppressed (Fig.~\ref{Fig:Fig_phase}~(a)).

At low packing fraction $\Phi$ (with poor clustering) $g(r)$ displays a first narrow and high peak before approaching one (Fig.~\ref{Fig:Fig_phase}~(c), (g), (k)).
This peak has no passive equivalent, i{.}e{.} at these values of $\Phi$, a passive system's $g(r)$ is flat.
Indeed, a dilute active system is characterized by numerous short-lived pairs with no passive counterpart.
The analysis of $g(r)$ for further values of the packing fraction $\Phi$ provides additional information on the cluster's internal structure. As expected, for increasing $\Phi$, $g(r)$ displays a second peak (Fig.~\ref{Fig:Fig_phase}~(d), (h), (l)), a third peak (Fig.~\ref{Fig:Fig_phase}~(e), (i), (m)) and a fourth peak (Fig.~\ref{Fig:Fig_phase}~(f), (j), (n)).
These peaks are due to the large size of the clusters that on average consist of more and more particle shells.
We identify the transition from liquid-like to solid-like structures when the second peak of $g(r)$ splits into two peaks.

When reduced inertia is high (small $f$), all the peaks of $g(r)$ are less pronounced and broader for all packing fractions.
This implies that the decrease in the shaker frequency affects the structural properties of the clusters, making them less cohesive.
In particular, this occurs in the configurations showing motility-induced clustering (see SM, Sec.~S2).
While these clusters are characterized by solid order (split second peak of $g(r)$) as in Refs.~\cite{digregorio2018full, caprini2020hidden}, here we prove the existence of cluster states with a liquid-like structure, i{.}e{.} ones characterized by a $g(r)$ with a single second peak.
This conclusion is additionally supported by the plots of the $\Psi_6$-map in experimental configurations (see Appendix~\ref{sec:psi6}, Eq.~\eqref{eq:Psi6}, for the $\Psi_6$ definition), showing typical clusters with liquid-like (Fig.~\ref{Fig:Fig_phase}~(o)) and solid-like (Fig.~\ref{Fig:Fig_phase}~(p)) structures.
 Systematic evidence of the structural change for different shaker frequencies and packing fractions is shown and further discussed in SM, Sec.~S2.
This effect is induced by the increase of reduced inertia.
Large inertia implies more bouncing events, which hinder the particles' ability to remain stuck in highly packed structures with hexagonal order. 
Results are summarized in a phase diagram that distinguishes between the three states identified (Fig.~\ref{Fig:Fig_phase}~(b)): homogeneous phase at low packing fraction and large reduced inertia (yellow circles); motility-induced clustering characterized by liquid-like order, induced by the inertia increase (red pentagons); usual motility-induced clustering with solid order for a large packing fraction and small reduced inertia (blue pentagons).

Our experimental results are constrained to finite size effects and to the presence of boundaries.
To support the validity of our conclusions beyond these limitations, we perform large-size numerical simulations by evolving $N=10^4$ interacting inertial active Brownian particles with repulsive interactions. To focus on the minimal ingredient leading to the solid-liquid transition in the cluster internal structure, we consider elastic collisions since the restitution coefficient estimated in experiments is large, $e\approx 0.9$.
Particles move persistently with inertial dynamics, which dissipates energy via a linear friction force, and are further subject to translational noise.
Additional details on the model, the particle dynamics, and the numerical integration scheme is reported in the Appendix~\ref{sec:simulations}.
In our simulations, we keep both the P\'eclet number and the packing fraction fixed, while we vary the particle reduced inertia $M=m/(\gamma\tau)$ i.e.\ the ratio between the damping time and the persistence time. In this way, we explore the role of inertia at a fixed motility, clarifying the primary role of the former in this phenomenon.

For small inertia values, the main cluster numerically observed is characterized by the hexagonal order typical of two-dimensional crystals (Fig.~\ref{Fig:Fig_phase}~(q)), as revealed by plotting the heat map of $\Psi_6$ as a color gradient (see Appendix~\ref{sec:psi6}, Eq.~\eqref{eq:Psi6}, for the $\Psi_6$ definition). Correspondingly, the distribution $\text{Prob}(\Psi_6)$ in the cluster reveals a narrow peak close to $\Psi_6\sim 1$ and only a long, thin tail towards values smaller than 1 (Fig.~\ref{Fig:Fig_phase}~(t)). This tail is due to the defect chains present in the cluster as in Ref.~\cite{digregorio2022unified}.
As inertia increases, the cluster shows the coexistence of solid and liquid regions (Fig.~\ref{Fig:Fig_phase}~(r)), which reflects onto the bimodality of $\text{Prob}(\Psi_6)$ (Fig.~\ref{Fig:Fig_phase}~(u)).
Finally, a further inertia increase leads to a cluster with liquid structure (Fig.~\ref{Fig:Fig_phase}~(s)), with $\text{Prob}(\Psi_6)$ peaked at a value smaller than 1 as in liquid configurations (Fig.~\ref{Fig:Fig_phase}~(v)).
Only in this fully liquid case, $\text{Prob}(\Psi_6)$ calculated inside and outside the cluster are close, having both a liquid structure.

Therefore, the numerical analysis supports our experimental findings: inertia induces a solid-liquid change in the internal structure of clusters consisting of active particles.
Simulations show that these results are purely induced by inertia and do not depend on boundaries.
In addition, we remark that this effect is even stronger in our simulations where we have explored larger-size clusters compared to experiments to exclude finite-size effects.

\begin{figure*}[!t]
	\includegraphics[width=\textwidth]{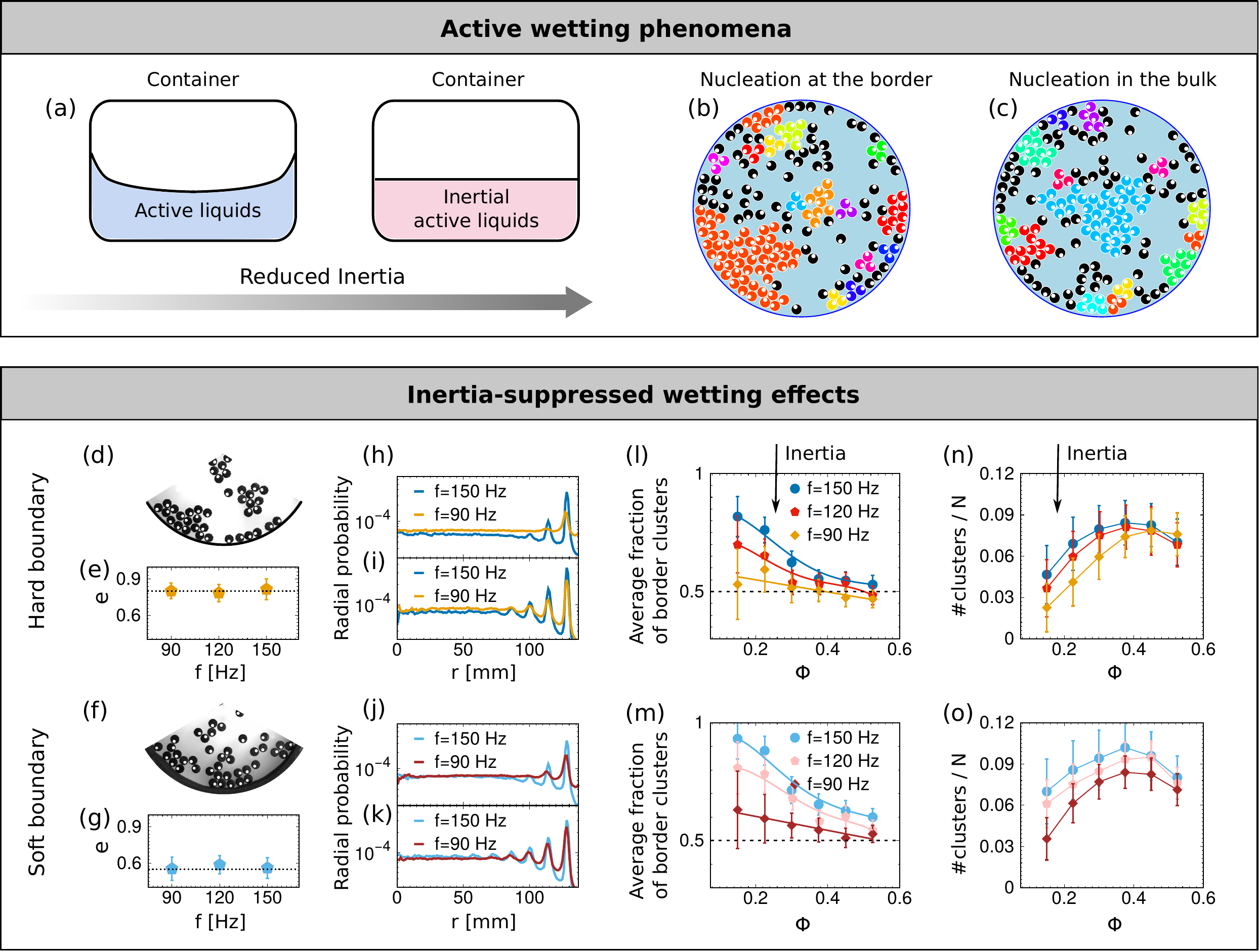}
	\caption{
		\label{Fig:Fig_Hetero}
		\textbf{Inertia-induced suppression of wetting phenomena}.
		(a)~Illustrations of wetting effects for active liquids and inertial active liquids in a container.
		(b)-(c)~Experimental snapshots showing cluster nucleation at the border and in the bulk of the container, respectively, for packing fraction $\Phi=0.45$ at shaker frequency $f=\unit[150]{Hz}$ (\textbf a) and $f=\unit[90]{Hz}$ (b). 
		(d), (f)~Portion of experimental snapshots for hard plastic (\textbf d) and soft sponge boundary (f).  
		(e), (g)~Restitution coefficient $e$ as a function of $f$ for the particle-wall collision (see SM, Sec.~S1 for the definition of the restitution coefficient) for the hard (e) and soft (g) boundaries.
(h)-(k)~Radial probability distribution from the plate center to the plate border for packing fractions $\Phi=0.15$ (h),  (j) and $\Phi=0.3$ (i), (k) for $f=\unit[150]{Hz}, \unit[90]{Hz}$. Furthermore, (h)-(i) are obtained for the hard boundary, while (j)-(k) for the soft boundary.
		(l)-(m)~Average fraction of clusters at the border of the container as a function of the packing fraction $\Phi$ for different shaker frequencies $f$.
		(n)-(o)~Total number of clusters (close and far from the boundary) normalized by the number of particles $N$ as a function of $\Phi$ for different $f$.
		Error bars in (e), (g) and (l)-(o) represent the standard deviation. 
}
\end{figure*}

\subsection{Inertia-induced suppression of wetting phenomena}

In the absence of inertia, active particles display spontaneous wetting phenomena in the presence of boundaries or obstacles with no passive counterpart: Indeed, in passive systems, a wetting effect can be observed only in the presence of attractive walls.
In general, overdamped active fluids display wetting phenomena~\cite{turci2023partial} on hard, pure repulsive boundaries since particles tend to accumulate near walls~\cite{hill2007hydrodynamic, kudrolli2008swarming, vladescu2014filling, vutukuri2020active} because of their persistent motion.
This implies that, typically, walls and system boundaries favor cluster nucleation in overdamped active matter~\cite{yang2014aggregation}.
In systems showing swarming or flocking with particles aligning to the wall profile, flower-like geometries have been employed to decrease boundary accumulation~\cite{deseigne2012vibrated}, effectively suppressing undesired wetting effects. However, this strategy does not work for our cylindrical particles that are not characterized by self-alignment or explicit torque at the boundary.
Here, we show that wetting effects can be suppressed by inertia (Fig.~\ref{Fig:Fig_Hetero}~(a)).

Having the circular boundary in our experimental setup allows us to systematically compare the cluster formation at the walls and in the middle of the container (Fig.~\ref{Fig:Fig_Hetero}~(b)-(c)) as well as the radial probability distribution calculated from the plate center to the border (Fig.~\ref{Fig:Fig_Hetero}~(h)-(k)).
In this way, we directly monitor the effect of inertia on the wetting phenomena, by changing the shaker frequency $f$.
This study is performed by considering two distinct boundaries confining active particles on the plate: The first is plastic-made and can be considered a hard wall (Fig.~\ref{Fig:Fig_Hetero}~(d)), the second is sponge-made and hence softer (Fig.~\ref{Fig:Fig_Hetero}~(e)).
The two walls are characterized by different dissipation and friction properties that manifest during a particle-wall interaction. These properties are estimated by evaluating the restitution coefficient $e$ (see SM, Sec.~S1 for the definition of $e$) by monitoring the particle velocity before and after a collision. While $e$ does not change with the frequency $f$, we observe a strong decrease from hard to soft wall, as expected (Fig.~\ref{Fig:Fig_Hetero}~(e)-(g)).

Wetting phenomena are predominant for the strongly damped case (high shaker frequency), as usual in colloidal systems that show clustering at high density~\cite{zottl2016emergent}.
In particular, a significantly larger number of clusters is observed adjacent to the wall than in the interior of the container, as observed by monitoring the number of clusters touching the boundary as a function of the average packing fraction $\Phi$ (Fig.~\ref{Fig:Fig_Hetero}~(l)-(o)).
In general, this cluster number decreases as $\Phi$ is increased. Indeed, for large packing fractions, the system is more crowded and particles can form a dynamical cluster before having the time to approach the border of the container.
Even if this tendency is shown for both boundaries, wetting effects are stronger for the sponge-made wall (see SM, Sec.~S2), which is characterized by a smaller restitution coefficient.
Indeed, during a particle-wall collision, friction is larger for sponge materials than for hard-plastic ones.
This trend is particularly strong for small $\Phi$, when cluster nucleation primarily starts at the boundary.

In striking contrast, high reduced inertia strongly suppresses accumulation at walls and wetting effects for both boundaries, since there is no longer a dominance by wall-induced clusters in this regime.
The hampering of wall accumulation is evidenced by the radial probability distribution: its peak at the boundary position is lower when inertia is increased, as checked for different packing fraction values and for the two boundary materials (Fig.~\ref{Fig:Fig_Hetero}~(h)-(k)).
In addition, we observe the suppression of wetting effects due to inertia because the fraction of border clusters is always lower when the shaker frequency is decreased.
When this measurement approaches values close to $1/2$, as it does for large inertia, we conclude that the probabilities of cluster nucleation at the wall and in the bulk are similar.
This unusual behavior is a surprising consequence of the bounce-back effect caused by inertia.
Indeed, for a small inertia, active particles attached to the wall slide along the boundary, so that when they collide with each other they get stuck more easily while they would move apart in a typical collision in the bulk.
By contrast, at high effective inertia, active particles bounce at the walls. In this way, when they encounter each other they tend to remain less stuck than in a bulk collision.
The suppression of wetting behavior and wall accumulation is qualitatively observed for both boundary materials even in the sponge-made case where the effective trapping power of the wall is larger. As a consequence, such an effect qualitatively occurs independently of the boundary details, such as the wall-particle restitution coefficient, and thus can be considered a purely inertia-induced effect.

\section*{Discussion}
Here, we have shown experimental evidence for the role of inertia in active matter.
By considering active granular particles, we have shown an inertia-induced transition from a solid-like to a liquid-like cluster before dynamical clustering is suppressed. In addition, inertia annihilates spontaneous wetting phenomena typical of active matter, as investigated for hard and soft boundaries.
Moreover, we experimentally observe a kinetic temperature difference inside and outside the cluster (see SM, Sec.~S2), confirming previous numerical results \cite{mandal2019motility}. Such a phenomenon has no passive counterpart since coexistence phases in equilibrium have the same temperature.
These effects are possible because of the active matter's non-equilibrium nature since inertia cannot affect the phase diagram in passive equilibrium systems.

In passive granular particles clustering is usually due to inelastic collisions~\cite{kudrolli1997cluster, olafsen1998clustering, brilliantov2004transient, chen2023clusters}, and favored by long-range electrostatic forces~\cite{lee2015direct} and acoustic forces~\cite{lim2019cluster}.
In our system of active vibrobots, on the other hand, clustering is mainly induced by motility.
This offers a new tuning parameter for structure formation in particulate matter.

Our findings will also stimulate future theoretical research in active systems since including inertia in existing field theories~\cite{te2023microscopic} is required for a full thermodynamic description of active matter~\cite{dabelow2019irreversibility, pietzonka2019autonomous, bowick2022symmetry}. We demonstrate how the novel effects observed here represent a promising research line to deepen the theoretical understanding of inertia-induced phenomena.

Our experiments show that, for vibrobots, inertia can be easily tuned by modifying the driving frequency.
These results can be pivotal in designing smart granular materials, consisting of units whose self-aggregation and structural properties can be externally controlled and manipulated to enhance cohesiveness and tune cluster speed.
Our findings could be further crucial to design materials useful in industrial applications to prevent agglomeration during mixing or to reach the desired granule size~\cite{trunec2014advanced}.

\section*{Acknowledgments} 
LC acknowledges support from the Alexander Von Humboldt foundation and acknowledges the European Union MSCA-IF fellowship for funding the project CHIAGRAM. HL acknowledges support by the Deutsche Forschungsgemeinschaft (DFG) through the SPP 2265, under grant number LO 418/25-1.

\section*{Author contributions}
LC and HL proposed the research.
LC, AL, and CS designed and manufactured the experimental setup.
LC carried out the experiments. LC and DB analyzed the experimental data, while LC performed numerical simulations.
All authors discussed the results and contributed to writing the manuscript.\par

\appendix


\section{Experimental details}\label{sec:experimental}

\noindent
{\bf{Description of the experimental setup.}}
The setup of our experiment consists of a standard vibrating table, commonly used for vibrational excitations of granular particles~\cite{Scholz2018inertial}. The enclosure for the particles consists of a cylindrical acrylic baseplate with diameter $\unit[300]{mm}$ and height $\unit[15]{mm}$ and an outer plastic ring that confines the particles.
The plate is fixed to an electromagnetic shaker and adjusted horizontally to an accuracy of about $\unit[10^{-2}]{degrees}$. In this way, the effect of any gravitational drift becomes negligible.
To suppress resonance effects with the environment, the setup is placed on a large concrete block.

The outer ring prevents particles from jumping out of the table and plays the role of a circular boundary condition.
In previous studies, flower-like box designs characterized by convex and concave regions helped to reduce boundary accumulation.
In particular, active granular particles characterized by self-alignment or polar interactions tend to align with walls. As a consequence, this wall shape favors particle migration to the middle of the box~\cite{deseigne2012vibrated} since those active granular particles slide along the wall profile.
The absence of self-alignment and polar interactions in our vibrobots would not allow this strategy to be particularly efficient.

In our study, we have explored two circular boundaries consisting of different materials to explore the suppression of wetting phenomena due to inertia.
In particular, we have considered a hard plastic boundary and a sponge-made boundary.
The former behaves as a hard wall, while the latter has a degree of softness and provides additional friction on vibrobots which hinders the particle motion. 
When not explicitly mentioned, the plastic boundary has been used.

\begin{figure}
	\includegraphics[width=\columnwidth]{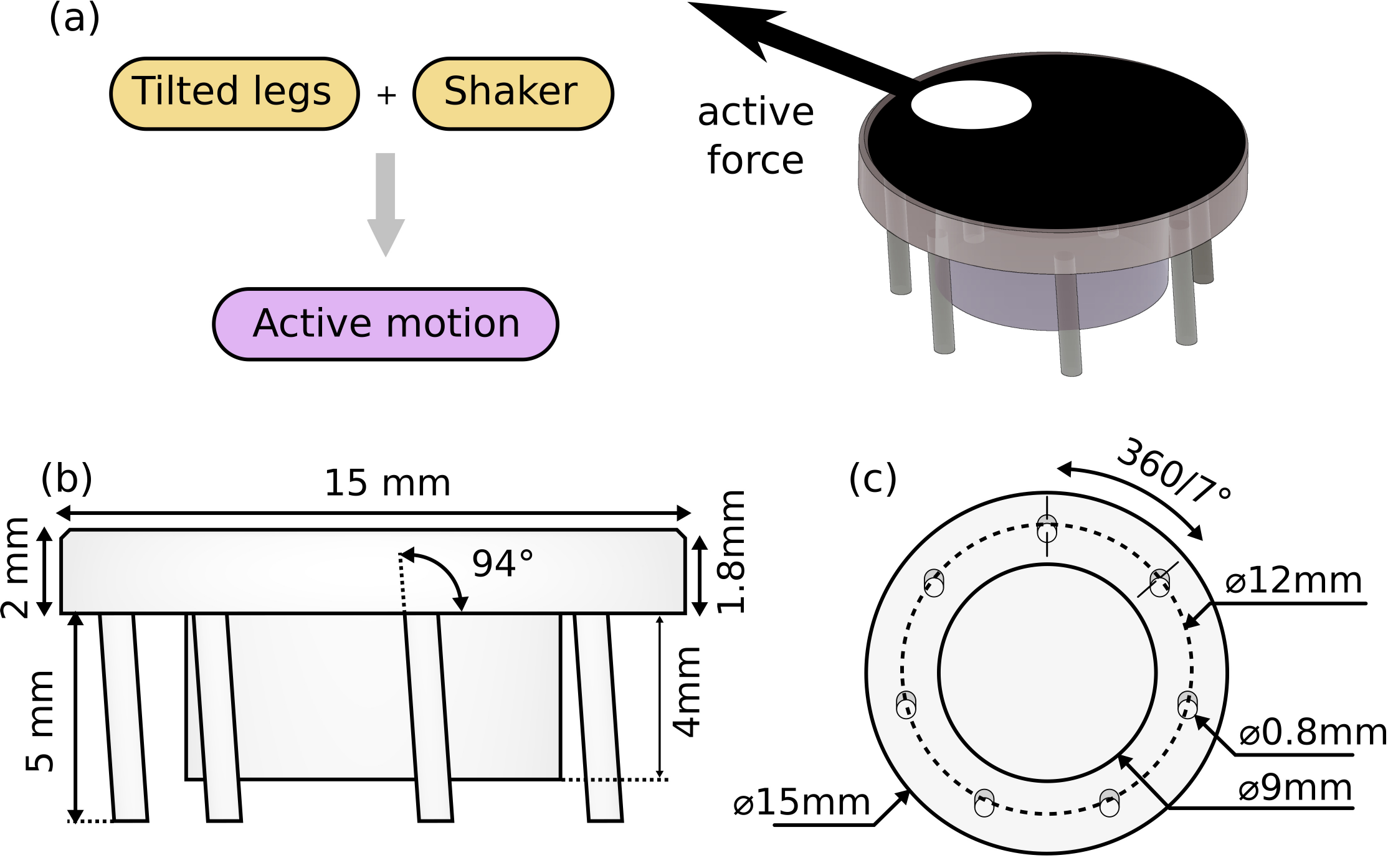}
	\caption{
		\label{Fig:Figscheme}
		\textbf{Particle shape}.
		(a)~3D illustration of the particles where the black arrow represents the direction of the active propulsion.
		(b)~Active vibrobot side-view schematic, showing typical sizes and the tilt angle.
		(c)~Active vibrobot bottom view schematic, showing the cylindrical symmetry uniquely broken by the tilted legs.
}
\end{figure}

\vskip0.2cm
\noindent
{\bf{Particle design.}}
The particle design follows the scheme introduced in~\cite{Scholz2018inertial}. The vibrobot body (see Fig.~\ref{Fig:Figscheme}~(a) for a 3D illustration) has a cylindrical core with diameter $\unit[9]{mm}$ and height $\unit[4]{mm}$. A larger cylindrical cap with diameter $\unit[15]{mm}$ and height $\unit[2]{mm}$ forms the top and defines the circular horizontal cross-section of the particle, such that the system can be considered quasi-two-dimensional. Seven tilted cylindrical legs with diameter $\unit[0.8]{mm}$ extrude from the bottom of the cap around the core (Fig.~\ref{Fig:Figscheme}~(b)), positioned on a regular heptagon at a distance of $\unit[6]{mm}$ from the center (Fig.~\ref{Fig:Figscheme}~(c)). The length is chosen such that the bottom face of the cap is $\unit[5]{mm}$ above the substrate and only the legs touch it.
To induce a directed motion the legs are tilted at a fixed angle $\delta=\unit[4]{degrees}$ relative to the surface normal.
On average the mass of the vibrobots is $m=\unit[0.83]{g}$.
The corresponding moment of inertia derived from the particle shape is $J=\unit[17.9]{g\,mm^2}$.

\vskip0.2cm
\noindent
{\bf{Shaker working regimes.}}
Frequency $f$ and amplitude $A$ of the shaker are chosen in suitable ranges to guarantee stable excitation, such that particles move without stopping but do not flip during the motion.
In particular, we have explored three different configurations: $f=\unit[150]{Hz}$ ($A=\unit[12(1)]{\upmu m}$), $f=\unit[120]{Hz}$ ($A=\unit[24(1)]{\upmu m}$), and $f=\unit[90]{Hz}$ ($A=\unit[45(3)]{\upmu m}$).

\begin{figure}
	\includegraphics[width=0.95\columnwidth]{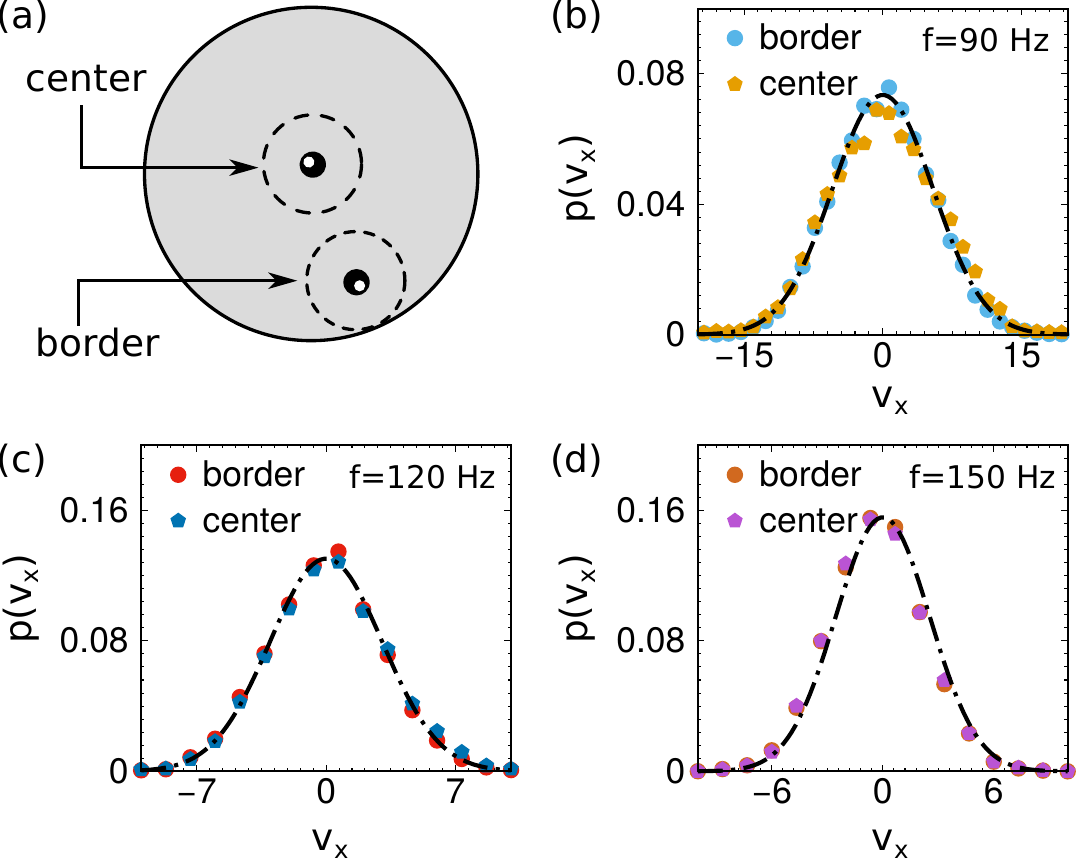}
	\caption{
		\label{Fig:Fig_spatial}
		\textbf{Negligible spatial inhomogeneity of the surface waves}.
(a)~Illustration of the experiment: passive particles are placed in the two virtual spatial regions marked by dashed black lines: i) center and ii) boundary.
(b)-(d)~Probability distribution of the particle velocity $p(v_x)$. Colored circles and pentagons referred to particles moving in the border and center regions, respectively. Black dashed-dotted lines are Gaussian fits of the experimental data. The study is performed for $f=\unit[90]{Hz}$ (b), $f=\unit[120]{Hz}$ (c), and $f=\unit[150]{Hz}$ (d).
}
\end{figure}

To guarantee that the change of shaker conditions leads to negligible spatially varying surface waves on the plate, we perform the following experiment:
We consider a passive particle with legs orthogonal to the plane of motion and a shape equal to active vibrobots. Being passive, these particles are more sensitive than active ones, and their use is more convenient to address this question.
The passive vibrobot is placed in two different circular regions on the plate (Fig.~\ref{Fig:Fig_spatial}~(a)): i) close to the center; ii) close to the boundary.
These virtual center and boundary regions are marked by black dashed lines in the figure.
When the particle diffusively approaches the contour of its region, the experiment is stopped and the particle is placed back in the middle of its circular region so that we separately sample only regions i) or ii). Compared to active vibrobots, passive particles move diffusively and thus they do not quickly escape the virtual circle allowing us to perform long experiments.
For the two spatial regions considered, we have plotted the velocity distribution $p(v_x)$, projected on a spatial component for $f=\unit[90]{Hz}$ (Fig.~\ref{Fig:Fig_spatial}~(b)), $f=\unit[120]{Hz}$ (Fig.~\ref{Fig:Fig_spatial}~(c)), and $f=\unit[150]{Hz}$ (Fig.~\ref{Fig:Fig_spatial}~(d)). For all the frequencies and regions explored, $p(v_x)$ has a Gaussian shape (dotted dashed line in Fig.~\ref{Fig:Fig_spatial}~(b)-(d)).
In addition, since in each graph the two $p(v_x)$ coincide, we conclude that the possible spatial inhomogeneity of surface waves plays a negligible role in the vibrobot properties.

\vskip0.2cm
\noindent
{\bf{Data acquisition and tracking.}}
In each configuration of density and frequency, we let the system evolve for 10 minutes to guarantee that the steady state is approached. Afterward, we acquire data for 20 minutes by recording the system with a high-speed camera at a resolution of $1024\times1024$ pixels and a frequency of $150$ frames per second. Vibrobot positions and orientations are detected using standard feature recognition methods (circle Hough transform) with a custom classical algorithm for sub-pixel accurate localization. From the particle positions in individual frames, we reconstruct the particle trajectories by identifying the nearest neighbors between a pseudo-frame extrapolated from the last known frame and velocity and the current frame. Translational and angular velocities are then calculated from the displacements as $\mathbf{v}=(\mathbf{x}(t+\Delta t)-\mathbf{x}(t-\Delta t)/(2\Delta t)$ and $\mathbf{\omega}=(\mathbf{\theta}(t+\Delta t)-\mathbf{\theta}(t-\Delta t))/(2\Delta t)$, with $\Delta t=\unit[0.0066]{s}$.
These velocities correspond to finite displacements, but due to the rather large inertial relaxation time of the particles, this is a sufficiently accurate approximation of their instantaneous velocities.

\section{Estimate of particle motion parameters}\label{sec:parameters}

To extract single-particle parameters, we perform single-particle experiments: We randomly place a particle on the plate, let the system evolve, and then track its trajectory until it reaches the boundary of the container. Afterwards, we manually reset the particle far from the boundary and repeat the procedure. This protocol excludes the slowing down due to other particles and the effect of the boundaries and allows us to extract parameters characterizing the single-particle motion.
For each frequency value, we consider 100 trajectories.

\begin{figure}
	\includegraphics[width=0.95\columnwidth]{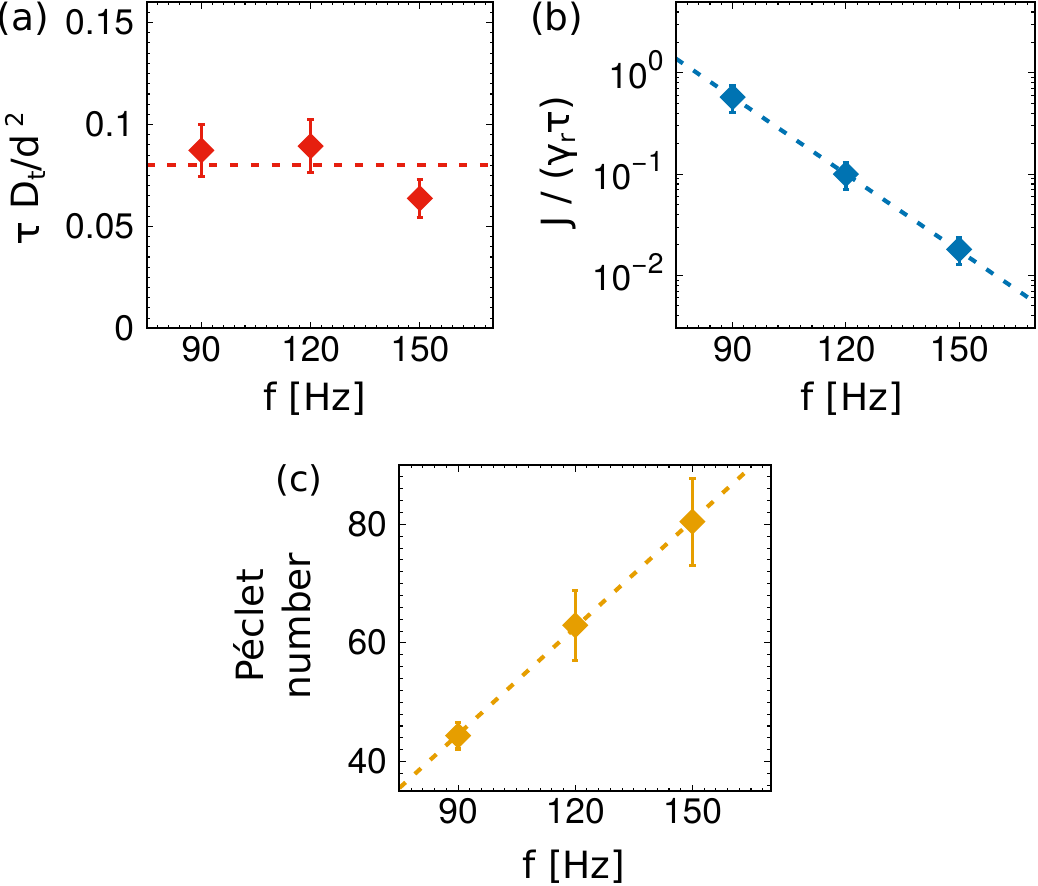}
	\caption{
		\label{Fig:Fig_dimensionless}
		\textbf{Other parameters of the single active particles}.
		(a)~Reduced translational diffusion coefficient, $\tau D_t/d^2$, as a function of the shaker frequency $f$.
		(b)~Reduced rotational inertia, $J/(\gamma_r\tau)$ as a function of $f$.
		(c)~P\'eclet number, $\tau v_0/d$, as a function of $f$.
		Points are obtained by the fitting of experimental data, error bars are the standard deviation, and dashed lines are eye guides. 
	}
\end{figure}

\vskip0.2cm
\noindent
{\bf{Fitting method to extract single-particle parameters.}}
The dynamics of particles at different frequencies is parametrized by modeling their motion with inertial active Brownian particle dynamics.
Our algorithm fits the dynamics to the measurements of several observables~\cite{Scholz2018inertial}, such as
translational and angular velocity distributions, the mean squared displacement and the mean angular displacement.
These observables are evaluated for an initial set of parameters, and the total deviation from the measured results is calculated and minimized by using a Nelder-Mead optimization scheme until an optimal set of parameters is found.

According to the inertial active Brownian particle model, the translational motion of active vibrobots with mass $m$ is described by a two-dimensional stochastic differential equation for position $\mathbf x$ and velocity $\mathbf v=\dot{\mathbf x}$
\begin{equation}\label{eq:dynamics_translational}
m\dot{\mathbf v}+\gamma\mathbf v=\gamma\sqrt{2D_\text{t}}\boldsymbol\xi+\gamma v_0\mathbf n\,,
\end{equation}
which balances inertial, dissipative, and active forces.
$\boldsymbol\xi$ is Gaussian white noise with zero average and unit variance, $\gamma$ describes the friction coefficient, and $D_\text{t}$ represents the effective translational diffusion coefficient.
The active driving force is given by $\gamma v_0\mathbf n$ with active speed $v_0$ and orientation vector $\mathbf n=(\cos\theta,\sin\theta)$ in terms of orientational angle $\theta$.
A vibrobot with finite extent and moment of inertia $J$ is characterized by an additional equation of motion for the angular velocity $\omega=\dot\theta$
\begin{equation}\label{eq:dynamics_rotational}
J\dot\omega=-\gamma_\text{r}\omega+\gamma_\text{r}\sqrt{2D_\text{r}}\eta\,,
\end{equation}
where $\eta$ represents Gaussian white noise with zero average and unit variance, and $\gamma_\text{r}$ and $D_\text{r}$ denote the rotational friction and the rotational diffusion coefficients, respectively. The inverse of $D_r$ defines the active particle persistence time $\tau=1/D_r$.

\section{$\Psi_6$ definition}\label{sec:psi6}

In Fig.~\ref{Fig:Fig_phase}, we have shown that inertia is able to affect the phase diagram of active particles not only by hindering clustering but also by inducing a solid-liquid transition in the cluster's inner structure.
To distinguish solid-like and liquid-like clusters in Fig.~\ref{Fig:Fig_phase}, we have evaluated the pair correlation function $g(r)$ of the system and classified different structures using a well-established criterion: a configuration is defined as solid if the second peak of the $g(r)$ is split while is defined as liquid otherwise. 
To further visualize the structural properties of each configuration, we have investigated the spatial map of the local director field $\Psi_6$ usually considered to estimate the orientational order parameter and defined as
\begin{equation}
\label{eq:Psi6}
\Psi_6(\mathbf{r}_i)= \frac{1}{N_i}\sum_{j=0}^{N_i} e^{i 6 \alpha_{ij}} \,,
\end{equation}
where the sum runs over its number of nearest neighbors $N_i$ and $\alpha_{ij}$ represent the angles formed by the position of the $i$-th particle and the position of one of its nearest neighbors.
$\Psi_6$ is a complex number of magnitude $|\Psi_6 | \leq 1$ which phase provides the six-folded director.
As a consequence, $\Psi_6\sim1$ if the system is locally characterized by the hexagonal order typical of solid configurations in two dimensions and it assumes smaller values otherwise.

\section{Simulations of inertial active Brownian particles}\label{sec:simulations}

The inertia-induced solid-liquid-like transition in the cluster's inner structure is further investigated with numerical simulations based on a minimal model.
This numerical study helps to exclude that this result is a consequence of finite-size effects or is due to the presence of the boundary.

We consider $N$ interacting self-propelled particles, governed by the inertial active Brownian particle dynamics (inertial ABP).
Particle positions $\mathbf{x}_i$ and velocities $\mathbf{v}_i$ follow dynamics \eqref{eq:dynamics_translational} and are additionally subject to pure repulsive forces due to the other particles $\mathbf{F}_i$. The orientational angles $\theta_i$ and angular velocities $\omega_i$ are governed by non-interacting equations of motion \eqref{eq:dynamics_rotational}.
Therefore, the system is described by
\begin{subequations}\label{eq:interactingdynamics}
\begin{align}
&m\dot{\mathbf v}_i+\gamma\mathbf v_i=\gamma\sqrt{2D_\text{t}}\boldsymbol\xi_i+\gamma v_0\mathbf n_i + \mathbf{F}_i\\
&J\dot\omega_i=-\gamma_\text{r}\omega_i+\gamma_\text{r}\sqrt{2D_\text{r}}\eta_i\,,
\end{align}
\end{subequations}
with forces obtained from a potential $\mathbf{F}_i=-\nabla_i U_{tot}$.
$U_{tot}$ is pairwise such that $U_{tot}=\sum_{i<j} U(|\mathbf{x}_i -\mathbf{x}_j|)$ with $U(|\mathbf{x}_i -\mathbf{x}_j|)$ obtained from the Weeks-Chandler-Andersen potential
\begin{equation}
U(r)=4\epsilon \left( \left(\frac{d}{r}\right)^{12} - \left(\frac{d}{r}\right)^{6} \right)\,.
\end{equation}
Here, $d$ is the nominal particle diameter while $\epsilon$ is the potential energy scale.
Equations~\eqref{eq:interactingdynamics} do not include dissipative collisions in the dynamics. Indeed, the small value of the restitution coefficient, experimentally measured in our system, allows us to consider purely elastic collisions with a reasonable degree of approximation.
Additionally, the rotational dynamics does not involve self-alignment interactions or torques which are negligible for our vibrobots (see SM, Sec S3). 
We remark that these choices allow us to show that motility and inertia are the minimal ingredients to reproduce the cluster structure changes.

Simulations are performed by rescaling time with the persistence time $\tau=1/D_r$ and spatial coordinates with the particle diameter $d$.
In this way, the dynamics is governed by several dimensionless parameters. Three of them characterize the overdamped dynamics of active particles:
i) the P\'eclet number $\text{Pe}=v_0/(D_r d)$, namely the ratio between persistence length, $v_0/D_r$, and particle diameter.
ii) The reduced translational diffusion, $\tau D_t/d^2$, which quantifies the impact of the random translational noise on the dynamics.
iii) The interacting potential generates an additional parameter, i.e. the reduced potential energy scale $\sqrt{\epsilon/m}/(D_r d)$.
In addition, inertia induces two additional dimensionless parameters, as in the single-particle case, namely
iv) The reduced mass $M=\frac{D_r m}{\gamma}$, i.e. the ratio between inertial time $m/\gamma$ and persistence time $\tau=1/D_r$.
v) The reduced moment of inertia $\mathtt{I}=\frac{D_r J}{\gamma_r}$, i.e. the rotational inertial time $J/\gamma_r$ in units of persistence time $\tau=1/D_r$.
The dynamics~\eqref{eq:interactingdynamics} is integrated by using the Euler algorithm with time-step $10^{-5}\tau$.
Particles are placed in a square box of size $L$ with periodic boundary conditions at packing fraction $\Phi=0.45$ (with $N=10^4$ and $L=125.3\,d$).

To observe dynamical clustering and motility-induced phase separation at packing fraction $\Phi=0.45$ for small inertia, we set $\text{Pe}=50$, $\tau D_t/d^2=10^{-4}$, and $\sqrt{\epsilon/m}/(D_r d)=1$. Indeed, to observe clustering, it is necessary to choose a sufficiently large value of the P\'eclet number so that the flux of particles joining the cluster is larger than the flux of particles leaving the cluster because of the particle motility reorientation.
Apart from this condition, our parameter choice is irrelevant.
To evaluate the influence of inertia, we fix the reduced moment of inertia $\mathtt{I}=10^{-1}$ and vary the reduced mass $M$.
In Fig.~\ref{Fig:Fig_phase}~(q)-(s), we report snapshot configurations for three values of $M=10^{-2}$ (negligible inertia), $M=5 \times10^{-2}, 10^{-1}$ (relevant inertia) that show clustering. Larger values of $M$ lead to clustering suppression.
In each snapshot, we plot $|\Psi_6|$ as a color gradient by using its definition, i.e.~\eqref{eq:Psi6}.
For small inertia (Fig.~\ref{Fig:Fig_phase}~(q)), particles in the cluster are arranged in solid-like configurations with some internal defects.
The increase of inertia leads to a cluster characterized by the coexistence between solid and liquid phases (Fig.~\ref{Fig:Fig_phase}~(r)), while for further larger values of the reduced mass, the cluster displays liquid order (Fig.~\ref{Fig:Fig_phase}~(s)).

\subsection*{Data availability}
The data that support the plots within this paper and other findings of this study are available from the corresponding author upon request.\par

\subsection*{Code availability}
An STL file for the design active vibrobots is included as Supplemental Material.\par

\subsection*{Competing financial interests}
The authors declare no competing financial interests.\par

\bibliographystyle{naturemag}
\bibliography{bib}

\end{document}